\documentclass[reprint,aps,prx,letterpaper,longbibliography]{revtex4-2}
\usepackage{amssymb}
\usepackage{amsmath}

\usepackage{float}
\makeatletter
\let\newfloat\newfloat@ltx
\makeatother

\usepackage{algorithm}
\usepackage{algpseudocode}
\usepackage{graphicx}
\usepackage{footnote}
\usepackage{subcaption}
\usepackage[usenames,dvipsnames,svgnames,table]{xcolor}
\usepackage[T1]{fontenc}
\usepackage[
pdfauthor={Paul D. Nation, Matthew Treinish},
pdftitle={Suppressing quantum circuit errors due to system variability},
bookmarks=true,colorlinks=true,linkcolor=OrangeRed,urlcolor=RoyalBlue,citecolor=DarkOrchid]{hyperref}

\definecolor{lightgray}{gray}{0.95}
\let\oldtabular\tabular
\let\endoldtabular\endtabular
\renewenvironment{tabular}{\rowcolors{2}{white}{lightgray}\oldtabular}{\endoldtabular}

\begin{document}

\title{Suppressing quantum circuit errors due to system variability}

\author{Paul D. Nation}
\email[E-mail: ]{paul.nation@ibm.com}
\author{Matthew Treinish}
\affiliation{IBM Quantum, Yorktown Heights, NY 10598 USA}

\begin{abstract}
We present a quantum circuit optimization technique that takes into account the variability in error rates that is inherent across present day noisy quantum computing platforms.  This method can be run post qubit routing or post-compilation, and consists of computing isomorphic subgraphs to input circuits and scoring each using heuristic cost functions derived from system calibration data.  Using an independent standard algorithmic test suite we show that it is possible to recover on average nearly 40\% of missing fidelity using better qubit selection via efficient to compute cost functions.  We demonstrate additional performance gains by considering qubit placement over multiple quantum processors.  The overhead from these tools is minimal with respect to other compilation steps, such as qubit routing, as the number of qubits increases.  As such, our method can be used to find qubit mappings for problems at the scale of quantum advantage and beyond.
\end{abstract}
\date{\today}

\maketitle

\section{Introduction}\label{sec:intro}
Given the limitations of present day noisy quantum hardware, implementing quantum circuit error suppression techniques \cite{ding:2020, niu:2020, murali:2019, finigan:2018} is vital for obtaining high-fidelity results over non-trivial circuit space-time volumes \cite{sundaresan:2022, glick:2021, kim:2021, jurcevic:2020}.  Indeed, when compiling a quantum circuit to a given quantum system the choice of physical qubits, basis gates, swap mapping \cite{li:2018}, gate optimization \cite{vatan:2004}, and dynamical decoupling \cite{ezzell:2022, tripathi:2021, pokharel:2018} routines all play a role in helping to reduce errors when executing the circuit.

\begin{figure}[t]
\includegraphics[width=\columnwidth]{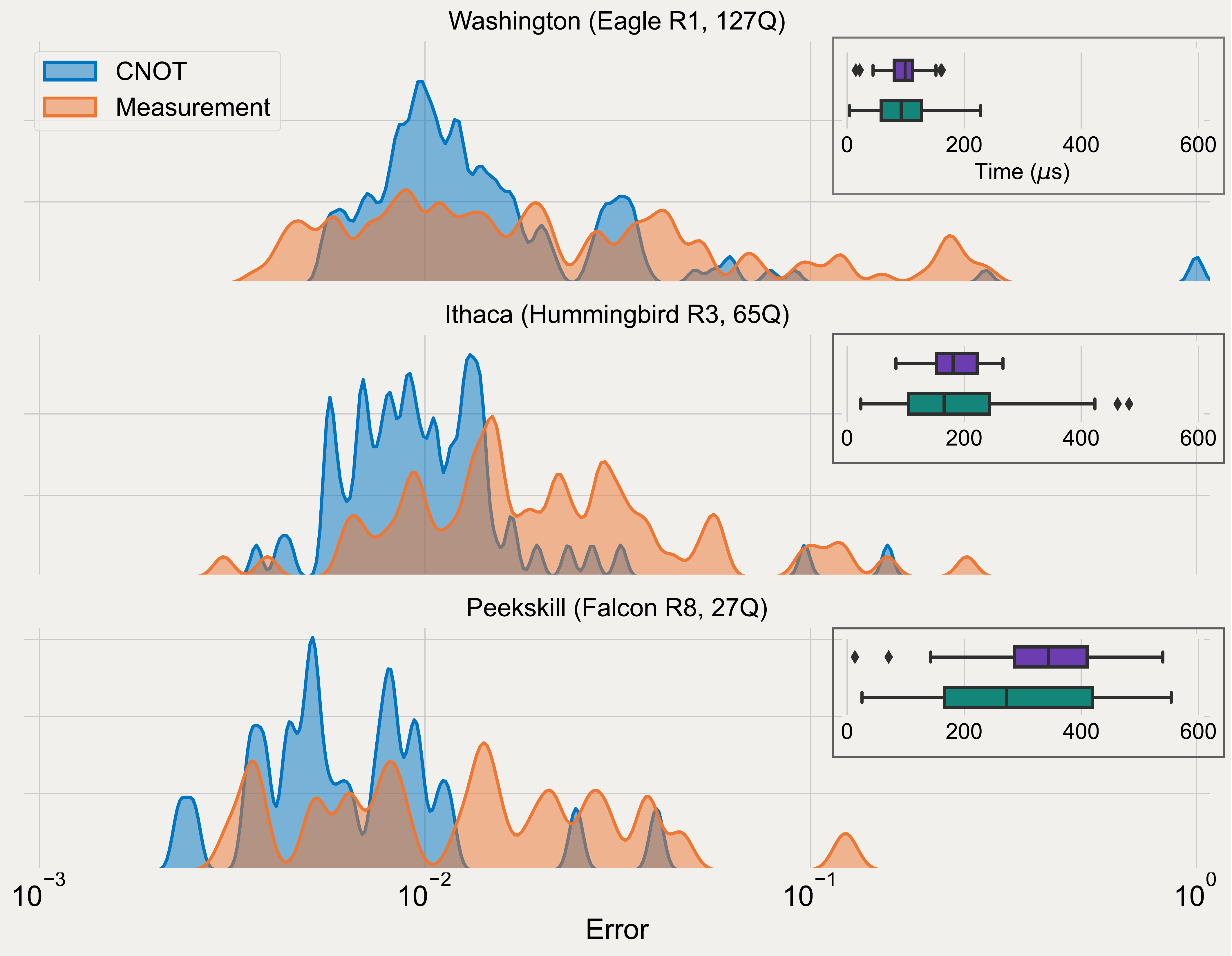}
\caption{Density of CNOT gate and measurement errors for three IBM Quantum systems representing different processor families, highlighting the variability of key performance metrics across the devices.  Insets show the distribution of $\rm T_{1}$ (top) and $\rm T_{2}$ (bottom) times.}
\label{fig:variability}
\end{figure}

Consideration of the noise characteristics of the target quantum system, e.g. gate and measurement errors and coherence times, is nominally taken into account \textit{at the beginning} of the compilation process via the choice of virtual to physical qubit mapping.  As shown in Fig.~(\ref{fig:variability}), present day quantum systems have substantial variation in their important metrics, making virtual to physical qubit mapping a crucial step in the compilation pipeline. The choice of qubits is even more critical when applying error mitigation techniques, where the mitigation process is exponentially sensitive on the fidelity of qubit operations \cite{berg:2022, nation:2021, temme:2017}.  

For small width circuits targeting hardware of $\mathcal{O}(10)$ qubits, it is usually possible to hand-select a low-noise subset of qubits with reasonable accuracy.  However, as hardware quality improves, and qubit counts begin to approach $\gtrsim \mathcal{O}(10^{2})$, finding optimal layouts manually becomes exceedingly difficult.  While automated noise-aware qubit placement routines do exist \cite{qiskit_dense_layout, murali:2019}, they often suffer from the fact that the best initial qubit layout in terms of noise characteristics is often not ideal for later qubit routing via swap mapping to the target topology; the error from additional swap gates dominates the savings from noise-aware qubit selection.  Moreover, the total number and type of gates in the final compiled circuit are not known ahead of time, further complicating qubit placement early in the compilation pipeline.  Additionally, it is possible to incorporate quantum processor noise into the qubit routing \cite{niu:2020} and approximate local gate compilation \cite{cross:2019}.  However these methods may lead to additional swap gates and poorly approximated global unitary operators, respectively.

Here we take a different approach, remapping quantum circuits after qubit routing, or post-compilation, to matching low-noise subgraphs using device calibration data.  Using output circuits generated with swap-efficient layout and routing routines that are not sensitive to device parameters, e.g. the Sabre method \cite{li:2018}, we compute the possible equivalent qubit mappings ranked by a heuristic cost function.  This can be done at the single device level, or across multiple machines of a similar topology.  Circuits are then remapped to the lowest error subgraph before execution.  A related technique was considered in Ref.~\cite{peters:2022} using Google Quantum AI processors \cite{google}, but it was found that the system calibration data failed to capture real-world performance details.  In marked contrast we will show via common independently defined algorithmic test circuits that meaningful improvements in quantum circuit fidelity can be obtained using even the most trivial of, and efficient to compute cost functions on IBM Quantum hardware.  We extend this technique across multiple quantum processors and highlight that additional gains can be found by relaxing the requirement that all circuits be executed on the same quantum system.  Although subgraph isomorphism is NP-complete \cite{10.1145/800157.805047}, we highlight that the cost of computing these graphs is much smaller than the cost of circuit compilation as a whole, and thus our method adds little in terms of relative cost.  Related machine learning algorithms have also been put forth \cite{cincio:2021}, but come with substantial overhead in the requirement of gate set tomography and circuit learning runtimes.

This paper is organized as follows.  In Sec.~(\ref{sec:method}) we detail how quantum processor subgraphs are determined from the entangling gate topology of a given quantum circuit and discuss how each subgraph is heuristically scored.  Section~\ref{sec:compiler} discusses implementation considerations when integrating our method into complete compilation pipelines.  In Sec.~(\ref{sec:benchmarking}) we look at the performance gains that are achievable with our technique using standard suites of algorithm test circuits, while Sec.~(\ref{sec:selection}) details additional performance gains that come when looking for optimal layouts across multiple quantum systems.  Finally, Sec.~(\ref{sec:conclusion}) summarizes our results and looks at possible future improvements.  Appendix~\ref{app:compare} looks at the effect of including $\rm T_{1}$ and $\rm T_{2}$ information in the cost function, while App.~\ref{app:mapping} details the importance of qubit routing when using our method. 

\section{Method}\label{sec:method}
Our algorithm, called \texttt{mapomatic} \cite{mapo}, is a post-routing routine that assumes that one or more circuits have been compiled to match the native gate set (basis gates) and entangling gate topology (coupling map) of a target device.  This guarantees that the graph structure of the circuits, as defined by their entangling gate connectivity, is a subgraph of the full device coupling map.  Finding an optimal set of qubits then becomes a two-step process.  First, a search over the system coupling map is performed to identify subgraphs that are isomorphic to each input circuit.  Second, a heuristic cost function is used to score the resultant mappings to find subgraphs with the lowest error.  Once identified, the input circuits are remapped to their corresponding minimal-cost subgraphs before execution.

\begin{figure}[t]
    \includegraphics[width=\columnwidth]{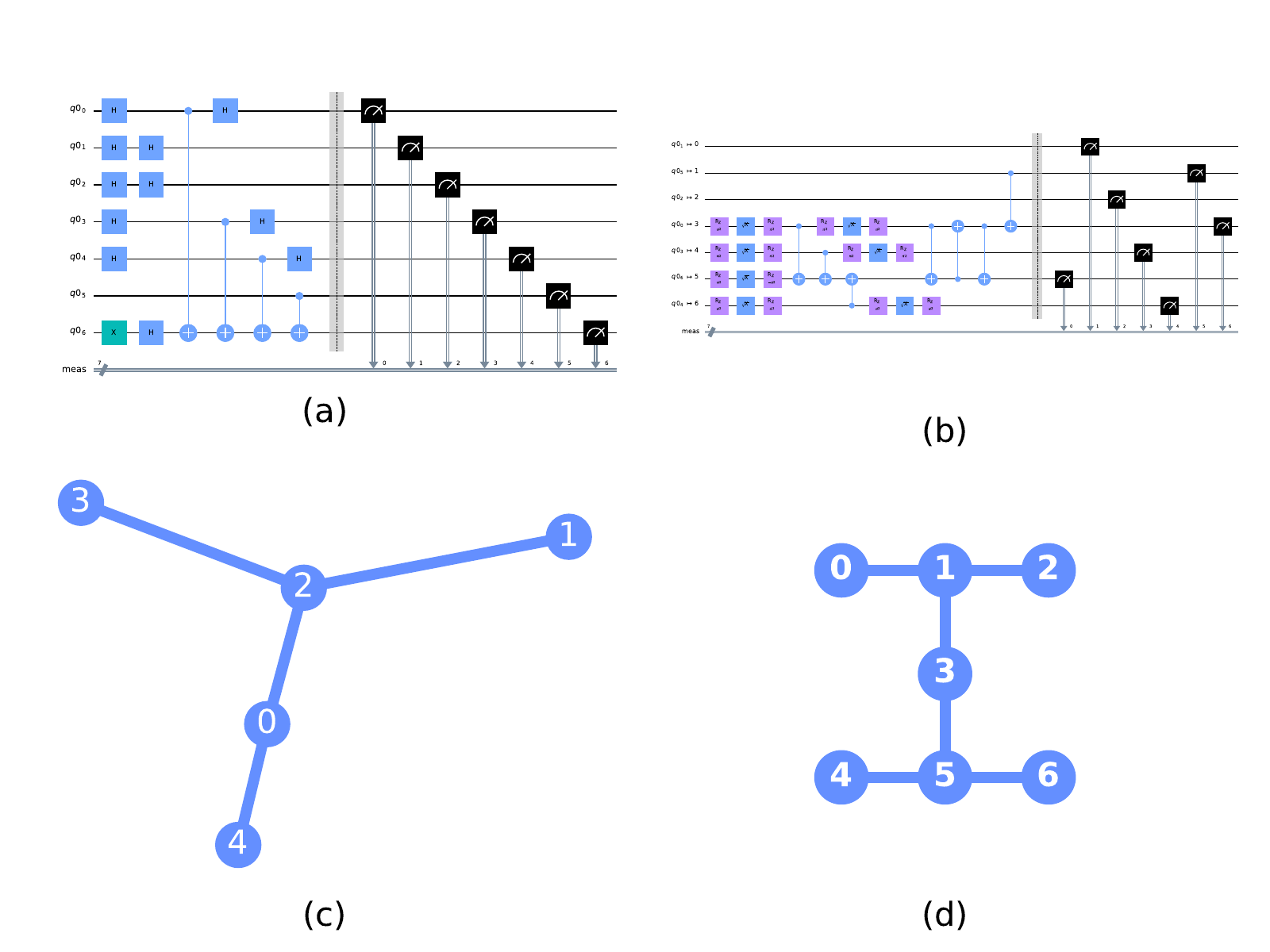}
    \caption{The stages of \texttt{mapomatic} algorithm. (a) Input circuit before compilation (b) Circuit after compilation targeting the IBM Quantum Nairobi system.  (c) Interaction graph for the compiled circuit. (d) Coupling map for the IBM Quantum Nairobi backend.}
    \label{fig:algorithm_overview}
\end{figure}

In step one, we iterate over all the instructions \footnote{Here, instructions include gates and non-unitary operations such as qubit reset and measurement, but not timing and alignment expressions such as barriers.}  in the
circuit after qubit routing, and possibly after additional gate optimization (post-compilation), and build an interaction graph. Edges in this graph represent unique two-qubit gates in the circuit, and single-qubit gates are treated as standalone nodes. Typically, a simple graph that is undirected and with no parallel edges is used.  We then search for isomorphic subgraphs on the quantum processor's connectivity graph; a graph where each node represents a physical qubit and each edge indicates support for two-qubit gates between those qubits. Subgraphs isomorphic to the original mapping are bijective layouts between virtual circuit qubits and physical qubits on the quantum processor.  Although here we assume all entangling gates are of the same type this is not a limitation of our routine. Figure \ref{fig:algorithm_overview}
shows an example of the graph construction used for the subgraph isomorphism problem.

\begin{figure}[t]
\includegraphics[width=\linewidth]{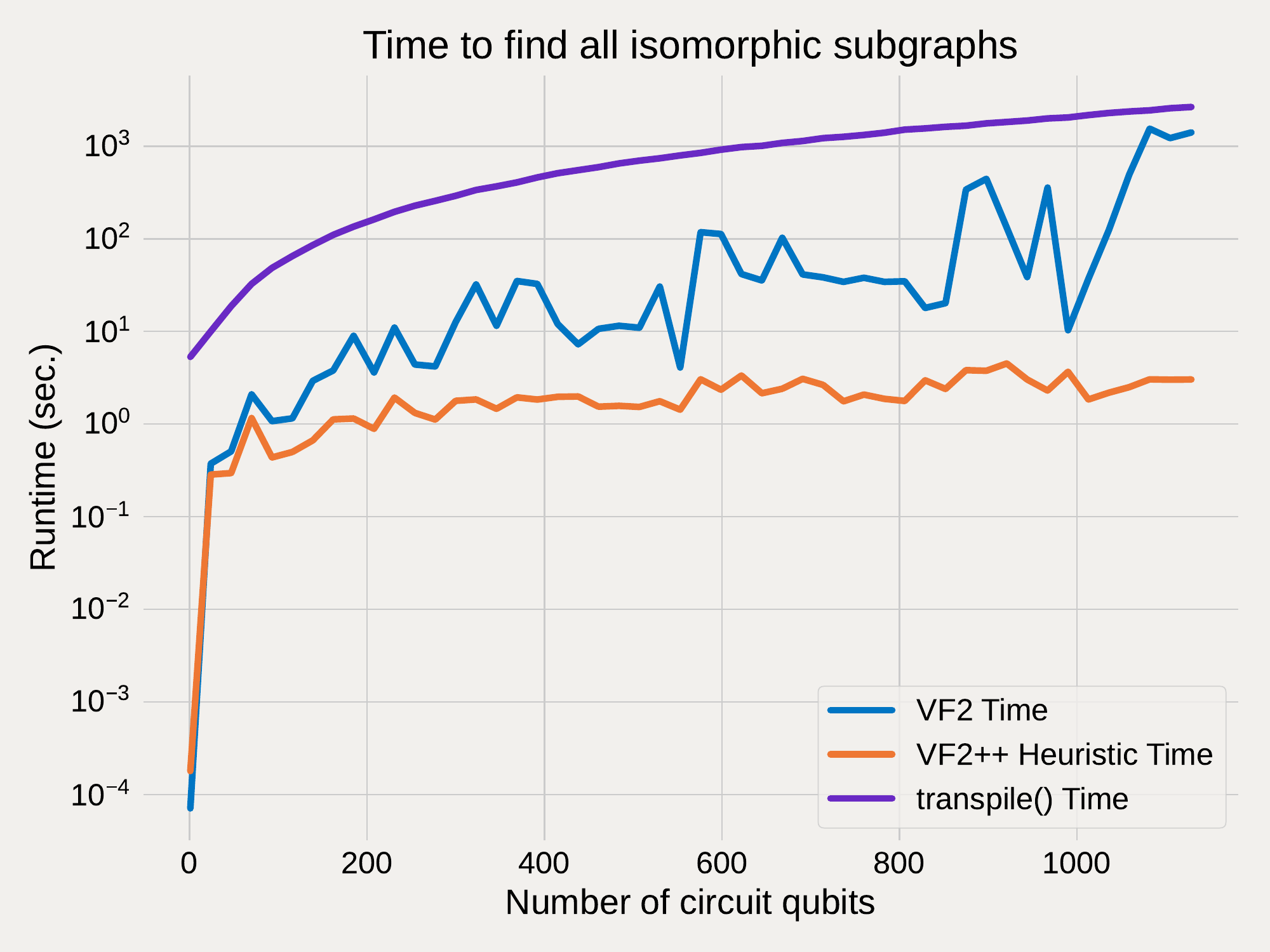}
\caption{Search time for finding all isomorphic subgraphs of an interaction graph of a
random $N$ qubit quantum circuit of fixed depth on the coupling map for a target backend with a 1299 qubit heavy-hex coupling map with VF2 and the VF2++ heuristic ordering. Time is compared to that of the Qiskit \texttt{transpile} function using the default \texttt{optimization\_level=1} on the same circuit. Benchmarks run on a AMD Threadripper 3970x with 128GB of RAM using Qiskit Terra 0.23.1 and rustworkx 0.12.1}
\label{fig:vf2_times}
\end{figure}

By running after the circuit has been mapped to system topology, there is at least one possible mapping available as the compiler must rewrite the circuit to match the device. In our implementation we use the \texttt{rustworkx} \cite{treinish:2021, retworkx} library's VF2 \cite{vf2} implementation which includes support for using the search order heuristic from VF2++ \cite{vf2++}. This node ordering heuristic, that \texttt{mapomatic} takes advantage of, orders the search-tree based on the degree of each node, and in some situations can make the isomorphism search
faster, minimizing the cost of the search. Figure \ref{fig:vf2_times} shows the run time of the \texttt{rustworkx} library's VF2 mapping function to find all the isomorphic subgraphs of the coupling map with and without the VF2++ ordering heuristic compared with the time it takes to execute the full Qiskit compilation pipeline without \texttt{mapomatic}.  Being three orders of magnitude faster at the largest qubit counts, it is clear that the overhead from subgraph finding is minimal compared to the rest of the compilation workflow.  This difference widens if, unlike the constant circuit depth used in Fig.~(\ref{fig:vf2_times}), we consider situations where the circuit depth increases with width.  In these cases, circuit routing and optimization time increases with depth, whereas subgraph finding is only minimally affected because the device topology remains unchanged.   Additionally, limitations can be set on the number of internal state visits used in VF2 to bound the overall runtime for finding a set of isomorphic mappings \footnote{The set need not be complete if the number of calls is not large enough.}.

For the second step we apply an efficient heuristic scoring function to each found mapping, Alg.~(\ref{alg:1}).
The scoring function bases its output on the reported calibration data for the processor from which we can define an error map $\mathcal E$ that maps physical instructions provided by a layout mapping $\mathcal M$ to error rates. For IBM Quantum systems this includes, amongst other information, single- and two-qubit gate errors, measurement infidelities, and $\rm T_{1}$ and $\rm T_{2}$ times across the device.  This data is nominally updated daily, and thus the scoring can change on similar timescales.  For each isomorphic mapping we estimate the overall fidelity of the circuit with that layout applied by taking the product of instruction fidelities corresponding to the physical qubits over which those instructions are applied.  That the resultant fidelity approaches zero in the large error limit is not a concern here as we are looking for relative differences, not absolute values [see Ref.~\cite{proctor:2022} for a discussion].
The returned scores for each layout are then used to rank all the possible mappings in
order of their estimated error, and the layout with the least error is used.  Note that the choice of cost function is not hardcoded into \texttt{mapomatic}, and users are free to define cost functions based on arbitrary input information.

\begin{algorithm}[b]
\caption{Algorithm for scoring layout mappings}
\label{alg:1}
\begin{algorithmic}[1]
\Procedure{Score}{$C, \mathcal{M}, \mathcal{E}$}
\Comment{Quantum circuit $C$ with a set of instructions $\{C\}$, layout mapping $\mathcal{M}$, and error map $\mathcal{E}$}
    \State $\rm fidelity \gets 1.0$
    	\For{$\rm instr \in \{C\}$}
        		\State $\rm fidelity \gets \rm fidelity * (1 - \mathcal{E}[\mathcal{M}[\rm instr]])$
   	 \EndFor
    \State \textbf{return} $1 - \rm fidelity$
\EndProcedure
\end{algorithmic}
\end{algorithm}

Missing from Alg.~(\ref{alg:1}) are $\rm T_{1}$ and $\rm T_{2}$ times, from which one can define approximate error rates associated with qubit idle times.  We do not include this information in our default cost function as it was empirically found not to have a large impact on the layout order from scoring; it nominally permutes qubits within a given mapping, but does not modify the mapping ordering, see App.~\ref{app:compare}. Additionally, adding timing information to quantum circuits is currently an unoptimized transformation in Qiskit \cite{qiskit}, and greatly increases the runtime of layout selection.  We do however include the cost function with idle errors as an example of a custom scoring function at the \texttt{mapomatic} website \cite{mapo}. 

It is also important to note that scoring returns floating-point values for each layout.  The difference between these values can be smaller than the uncertainty from device fluctuations, and ultimately finite-sampling statistics, effectively leading to a tie between one or more scored layouts.  In cases such as these, more complex cost functions and/or information beyond device calibration data is needed to break the scoring degeneracy.

\section{Integration into a compiler}\label{sec:compiler}

While \texttt{mapomatic} was originally designed to run as a standalone post-compilation routine it
is also possible to directly integrate it into a compilation pipeline. When doing this the only
additional constraint is to ensure that any changes in layout do not prevent the circuit from
running on the device. When running as a standalone post-compilation routine this is not a constraint
because you can always re-run the compilation with a fixed optimal layout to update the circuit. There
are two techniques for integrating the algorithm into a compiler, the first is to perform \texttt{mapomatic}
immediately after routing, the second is to run \texttt{mapomatic} after all optimization routines but prior to
any physical scheduling of the circuit. There are tradeoffs between the two approaches.

Running immediately after routing works with looser constraints typically ignoring directionality (by
using an undirected interaction graph and coupling map), and with no guarantee that the circuit has been
converted to the native gate set. This means that for error evaluation you can only apply an inexact
scoring, typically using average error rates for the different types of operations available (i.e. 1 qubit gates,
2 qubit gates, etc.) on the device instead of using the exact error rates for each instruction from the calibration
data. However, running with looser constraints provides the flexibility to evaluate more potential layouts,
and potentially yield better results as later compilation stages in the pipeline will be able to transform the
circuit as needed.

\begin{figure*}[t]
\includegraphics[width=\textwidth]{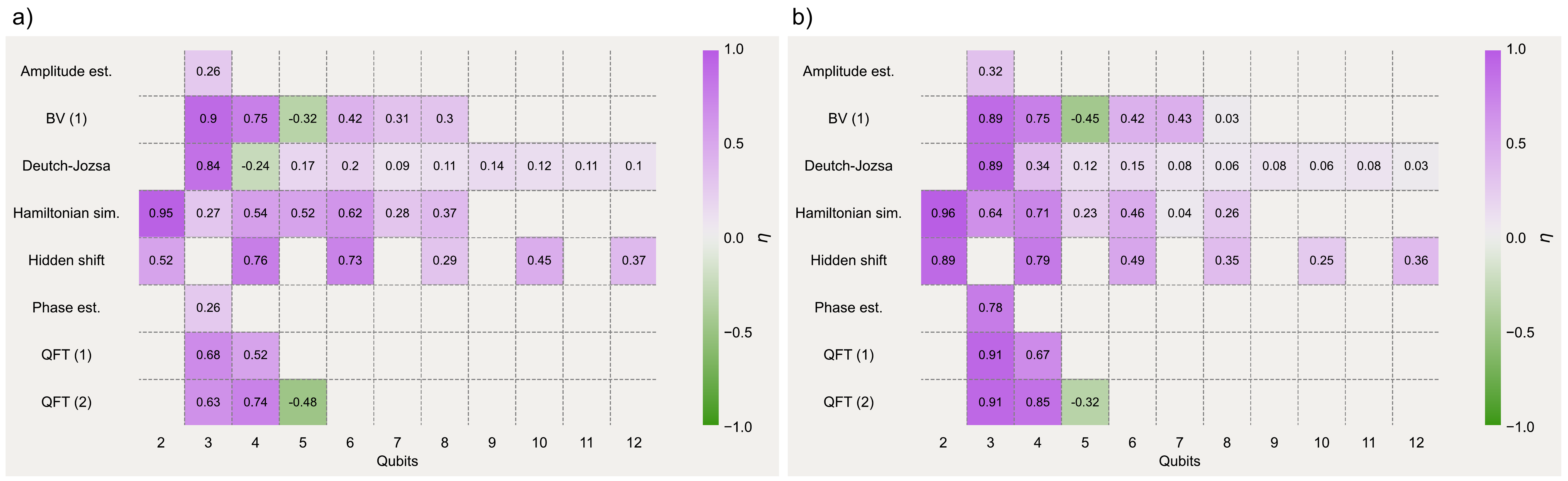}
\caption{Fraction of fidelity, Eq.~(\ref{eq:compare}), that is recoverable on the QED-C test circuits when using a) \texttt{mapomatic} to remap circuits generated by the Qiskit transpiler at optimal settings. b) Qiskit with \texttt{mapomatic} as a transpiler pass versus a baseline value corresponding to circuits compiled using PyTket.  Circuits are executed on the IBM Quantum Peekskill system taking $10,000$ samples with no gap between subsequent runs of the test suite.  Results colored purple show improvement over baseline, whereas green indicates degraded performance.  Results in (a) are produced using Qiskit Terra 0.20.2 with \texttt{mapomatic} where as (b) uses Qiskit Terra 0.21.0 in which \texttt{mapomatic} was introduced as a default transpiler pass.}
\label{fig:qedc1}
\end{figure*}

Running the algorithm after physical optimization (but before scheduling) means that the circuit
is guaranteed to be matching the directionality of the entangling gate topology and all operations are in
the native gate set. This allows the scoring Alg.~(\ref{alg:1}) to apply a more exact scoring as the
exact instructions used in the output circuit have already been determined leading to a potentially more accurate
selection of the best performing layout. However, running in this mode has more constraints on the
available layouts as at this point in a typical compilation pipeline the algorithm must conform to all the
constraints of the target device. This means the algorithm needs to use directed graphs for the interaction
graph and the coupling map and also only evaluate isomorphic mappings where the instructions performed on
the circuit qubits are available on the physical qubits selected. As selecting a layout that violates these
constraints would result in an invalid output from the compilation.

We've integrated the \texttt{mapomatic} algorithm into the Qiskit compiler as the \texttt{VF2PostLayout} pass \cite{qiskit-vf2postlayout}.
The \texttt{VF2PostLayout} pass supports running in both modes, however, it is integrated into the default compilation pipelines
only in the first mode that runs immediately after routing. While in most cases when compiling for current hardware
there is no difference when running the different techniques, there are some situations where
being able to evaluate more potential mappings can yield better results.

\section{Benchmarking results}\label{sec:benchmarking}
In order to validate the performance benefit of our method we will take advantage of the wide variety of test circuit libraries that are available \cite{tomesh:2022,lubinsky:2021,kohout:2020,li:2020}.  In particular, here we use circuits from the Quantum Economic Development Consortium (QED-C) \cite{lubinsky:2021}, comparing runs of this suite with and without \texttt{mapomatic}.  We point out that this test suite is developed independent of this work, and unlike a few hand selected examples, represents a true test of our technique.  Rather than focusing on all the tests results, some of which are well beyond what today's quantum processors are capable of, we only look at those algorithmic tests and number of qubits where at least one of the two fidelities used in the comparison is $\ge 1/e$.  This prevents inflated claims of success when dealing with differences in fidelity values whose overall magnitude is not indicative of meaningful experimental outcomes.  In addition, rather than looking at fidelities directly, we ask what fraction of the fidelity missing in the baseline result can be recovered through better qubit selection.  That is to say given a baseline fidelity $f_{\rm base}$ we compute the value

\begin{equation}\label{eq:compare}
\eta =\frac{f-f_{\rm base}}{1-f_{\rm base}},
\end{equation} 
where $f$ is the comparison fidelity.

In Fig.~(\ref{fig:qedc1}a) we use the fidelities obtained on the QED-C test suite using Qiskit as the baseline, and look at the fraction of recoverable fidelity when using \texttt{mapomatic} after compilation to remap the same circuit on the IBM Quantum Peekskill device.  Specifically, all baseline circuits were transpiled once in Qiskit using \texttt{optimization\_level=3} and \texttt{approximation\_degree=0} that enables Sabre layout and routing and disables approximate unitary synthesis.  We see that qubit remapping provides a substantial performance benefit over the qubit assignment of Sabre layout.  Across the benchmarking results, an average (median) of 37\% (34\%) of the fidelity is recoverable on the system via simple qubit remapping.  This emphasizes the importance of not only choosing the correct quantum system to execute circuits, but also the correct qubits on that system.  The small number of test results with negative values indicate a loss in fidelity from the remapping process; the calibration data does not faithfully represent a subset of qubits any more (device parameter drift) and/or that the simple cost function used here does not capture enough of the underlying contributions to circuit errors.  However, it is clear that in the vast majority of cases there is marked performance gains from using our technique to remap quantum circuits.  

While Fig.~(\ref{fig:qedc1}a) highlights fidelity improvements versus Qiskit without \texttt{mapomatic}, it is also beneficial to look at comparisons against other compilation pipelines.  To this end, in Fig.~(\ref{fig:qedc1}b) we look the differences in fidelity when using Qiskit with \texttt{mapomatic} integrated as a transpiler pass versus the compiler in PyTket \cite{sivarajah:2020} used as the base fidelity.  In this investigation we used PyTket version $1.3.0$ with \texttt{default\_compilation\_pass(2)}.  This comparison is of interest not only because PyTket is a commonly used alternative to Qiskit, but also because its qubit layout and routing methods are deterministic rather than stochastic.  As with Fig.~(\ref{fig:qedc1}a), we see a sizable portion of the fidelity missing from the PyTket results is recoverable with \texttt{mapomatic} utilized in the Qiskit compilation stack.

Although the results presented in Fig.~(\ref{fig:qedc1}) highlight the possible gains when using \texttt{mapomatic}, it is not a panacea, and other steps of the compilation process still play an important role in the final output fidelity.  In particular, the stochastic nature of the qubit routing (swap mapping) methods used in Qiskit gives rise to a variable number of SWAP gates in the final compiled circuits.  The variance on the number of added gates means that it is possible that a poor routing choice cannot be remapped with a higher fidelity than would otherwise be possible with a more careful routing selection utilizing repeated routing attempts, or using a non-stochastic routing routine such as that found in PyTket. An example highlighting poor routing on the overall fidelity is shown in App.~(\ref{app:mapping}).  As a corollary, our results show the challenge when evaluating the performance of quantum systems; the fidelity of final results depends greatly on the circuit rewriting pipeline used before execution.  This applies even more so to cross-platform comparisons where different compilation workflows, both client-side and server-side, further complicate the interpretation.

\section{Best device selection}\label{sec:selection}

To date, standard workflows for quantum computation involve first selecting a good target system on which to compile and execute a set of quantum circuits.  Until now, we have followed this procedure in this work as well, choosing the IBM Quantum Peekskill system based on prior knowledge of its performance characteristics.  However, as the number of available quantum systems grows, and the complexity of algorithms that they faithfully execute increases, it becomes challenging to ascertain which of the myriad of system and qubit layout combinations are good candidates.  Moreover, when running many different algorithms, or the same algorithm over a varying number of qubits, the optimal system and qubit layout can span several devices.  Therefore, it can be valuable to look for optimal circuit layouts across multiple quantum processors.

It is possible to use the tools presented here for determining good initial device and layout candidates across multiple quantum systems in the following manner.  To begin, circuits must be compiled against one of the systems within the set of possible target processors.  To obtain the largest number of matching subgraphs, it is ideal for the systems to share a common entangling gate topology, e.g. all IBM Quantum systems are based on the heavy-hex architecture \cite{chamberland:2020}.  The \texttt{mapomatic} routine can then be run across the set of quantum processors returning the device name, best layout, and the associated error value for each processor.  The device and layout corresponding to the lowest overall error value is then selected for remapping and execution.  The matching subgraphs for systems with the same topology need only be computed once, where as the cost function for each layout must be evaluated on each individual system.  If a quantum system does not have enough qubits to accommodate a circuit, then there are no matching subgraphs and the device is skipped. 

\begin{figure}[b]
\includegraphics[width=\linewidth]{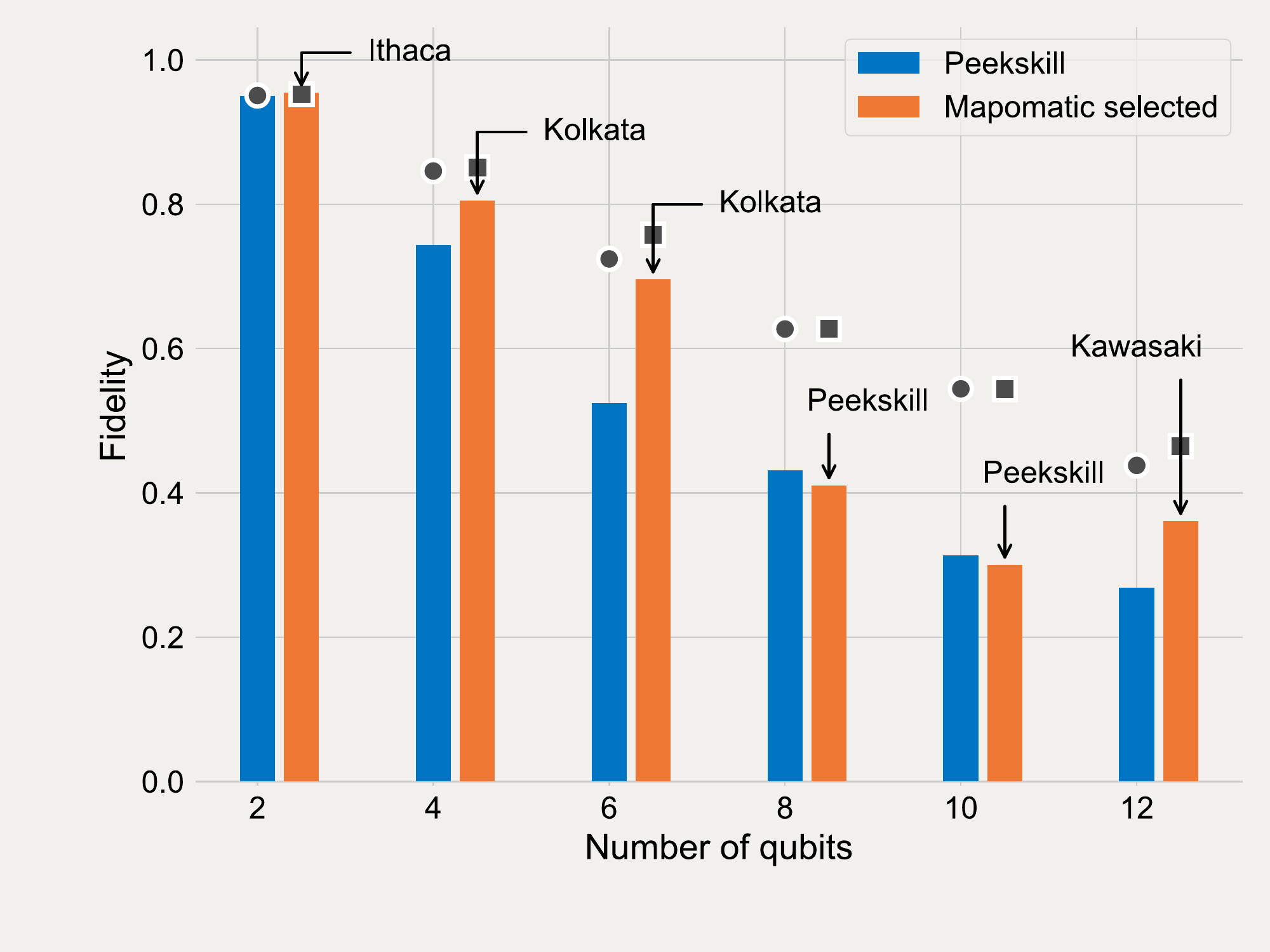}
\caption{Fidelity of QED-C Hamiltonian simulation circuits for the Peekskill system used in Fig.~(\ref{fig:qedc1}) compared to execution of the same circuits on the optimal target system as selected by \texttt{mapomatic} from the entire set of IBM Quantum devices.  Markers show the estimated fidelity for the Peekskill (circle) and \texttt{mapomatic} chosen (square) systems.  Circuits are executed in the same manner as Fig.~(\ref{fig:qedc1}).}
\label{fig:hamsim}
\end{figure}

To demonstrate the value in looking across multiple systems, we repeat the Hamiltonian simulation test circuits from the QED-C suite, and use \texttt{mapomatic} to find the system and layout with minimal cost across the entire fleet of IBM Quantum systems ranging from 5- to 127-qubits.  The experiments presented in Fig.~(\ref{fig:hamsim}) show the resultant fidelity from the quantum system selected from the entire IBM Quantum processor lineup compared to the best layout on the IBM Quantum Peekskill
system used in Fig.~(\ref{fig:qedc1}).  Although the Peekskill system is a newer high-coherence (avg. $\rm T_{1}\sim 300~\mu\rm sec$) 27-qubit Falcon R8 system, see Fig.~(\ref{fig:variability}), we see that previous generation Kolkata and Kawasaki Falcon R5 systems are selected and give better results for several circuit widths.  The same is true for the two-qubit case where the 65-qubit Ithaca system does better, but with gains limited by the overall high-fidelities at such small circuit space-time volumes.  We also see that the estimated fidelities returned by \texttt{mapomatic} do not match the results from hardware execution.  Firstly, as mentioned previously, we do not include qubit idle time errors in the cost function.  Although they contribute to the overall circuit error, we have empirically found that they do not greatly change the ordering of layouts; contributions from $\rm T_{1}$ and $\rm T_{2}$ in the instruction errors already capture much of much of these effects in the scoring process.  Second, the calibration data returned does not include sources of error such as spectator qubits or cross-talk.  Thus the fidelities corresponding to scored layouts should be loosely interpreted as upper-bounds.  The performance gains shown in Fig.~(\ref{fig:hamsim}) are non-negligible given that the Peekskill system is one of the best performing IBM Quantum systems to date, and was selected because of this.  This highlights that even with detailed device knowledge, looking across multiple quantum systems with \texttt{mapomatic} can provide performance gains that would otherwise be overlooked.

\section{Conclusion}\label{sec:conclusion}

We have shown that it is possible to account for system variability in near-term quantum processors using circuit remapping applied post-compilation, or after qubit routing.  Using simple, quick to evaluate cost functions we have demonstrated that sizable fractions ($\sim 40\%$) of result fidelity can be recovered on a wide variety of standard quantum application test circuits, and as compared to other circuit transformation pipelines, using our \texttt{mapomatic} method.  Using performant subgraph isomorphism routines, our technique adds little in terms of additional cost to the overall compilation process.  Given that much of the overall wall clock time of executing circuits is waiting in a queue, this remapping overhead can likely be amortized over this duration. This low-overhead also allows for layout scoring across multiple quantum processors, alleviating the need for users to identify a target system ahead of time, and uncovering additional performance improvements that would be missed if device selection is done ahead of time.  \texttt{mapomatic} is easy to integrate into quantum workflows and, because of the performance advantages it offers, is incorporated into the transpilation process by default in Qiskit Terra 0.21+.  Thus users are capable of leveraging these techniques immediately, and in many cases have been doing so without knowing it.

There are several possible future improvements to \texttt{mapomatic} that should be investigated.  Chief among them is looking for improved heuristics for scoring layouts that are more accurate while at the same time being efficient to evaluate.  In particular, is there information outside of standard device calibration data that can yield more accurate cost analysis and break ties between layouts?  Additionally, qubit selection over multiple quantum systems has received little attention to date, but is likely beneficial for algorithms that can be executed in parallel over multiple systems, for example variational algorithms \cite{bharti:2022,cerezo:2021} and circuit cutting \cite{peng:2020,bravyi:2016}. More broadly, as the performance improvements seen with \texttt{mapomatic} apply across the board, it is possible to integrate it into cloud-based workflows that would allow for the abstraction of device selection away from users.  This would not only lead to performance gains, but also removes the need for users not interested in device characterization to understand detailed system information.  

As the field of quantum computation approaches the frontier of Quantum Advantage, and users increasingly make use of error mitigation techniques with runtimes exponentially sensitive on qubit quality, \texttt{mapomatic} and related qubit selection methods will undoubtedly play a pivotal role in early demonstrations of practical quantum applications.

\begin{acknowledgments}
We thank Doug McClure and David McKay for helpful discussions.  This material is based upon work supported by the U.S. Department of Energy, Office of Science, National Quantum Information Science Research Centers, Co-design Center for Quantum Advantage (C2QA) under contract number DE-SC0012704
\end{acknowledgments}

\appendix

\section{Including $\rm T_{1}$ and $\rm T_{2}$ in cost function}\label{app:compare}

The default cost function in \texttt{mapomatic} does not include errors on idle qubits due to incoherent errors from $\rm T_{1}$ and $\rm T_{2}$ processes.  As mentioned in the main text, this is due for two reasons.  First, it is currently inefficient to add repeatedly generate timing information in Qiskit, and doing so would greatly increase the runtime of the scoring process.  Second, even when this information is included, the resulting qubit layouts are to large extent nominally simple permutations of the layouts computed without these error processes; effects from $\rm T_{1}$ and $T_{2}$ are, to large extent, already accounted for in the gate and measurement errors included in the device calibration data.  In Table~\ref{tab:compare} we give an explicit example of this, looking at the resulting optimal qubit layouts generated by \texttt{mapomatic}, both with and without idle-errors included in the cost function, from the Hamiltonian simulation test circuits from Sec.~\ref{sec:selection}.

\begin{table}[t]
    \begin{subtable}[h!]{0.5\textwidth}
    \centering
    \begin{tabular}{|c|c|c|c|}
\hline
Num. qubits & Default cost function\\
\hline
2 & $[5, 8]$  \\
\hline
4 & $[9, 8, 5, 3]$\\
\hline
6 & $[16, 14, 11, 8, 5, 3]$ \\
\hline 
8 & $[2, 3, 5, 8, 11, 12, 13, 14]$\\
\hline 
10 & $[2, 3, 5, 7, 8, 10, 11, 12, 13, 14]$ \\
\hline
12 & $[23, 21, 18, 15, 12, 11, 14, 13, 8, 5, 3, 2]$ \\
\hline  
\end{tabular}
\vspace{-0.05in}
\end{subtable}
    \begin{subtable}[h!]{0.5\textwidth}
    \centering 
    \vspace{0.1in}
    \begin{tabular}{|c|c|c|}
\hline
Num. qubits & Cost function with $T_{1}$ and $T_{2}$\\
\hline
2 & $[5, 8]$  \\
\hline
4 &  $[9, 8, 5, 3]$ \\
\hline
6 & $[16, 14, 11, 8, 5, 3]$ \\
\hline 
8 &  $[12, 13, 14, 11, 8, 2, 3, 5]$\\
\hline 
10 & $[3, 5, 8, 4, 11, 7, 14, 10, 12, 13]$ \\
\hline
12 & $[24, 23, 21, 18, 15, 14, 13, 12, 11, 8, 5, 3]$ \\
\hline  
\end{tabular}
    \vspace{-0.05in}
     \end{subtable}
    \caption{Lowest cost layouts found using the default \texttt{mapomatic} cost function,  as well as when including idle-errors due to $\rm T_{1}$ and $\rm T_{2}$ in the scoring process, for the Hamiltonian simulation experiments on IBM Peekskill presented in Sec.~\ref{sec:selection}.}
    \label{tab:compare}
\end{table}

\section{SWAP mapping variability}\label{app:mapping}

\begin{figure}[b]
\includegraphics[width=\columnwidth]{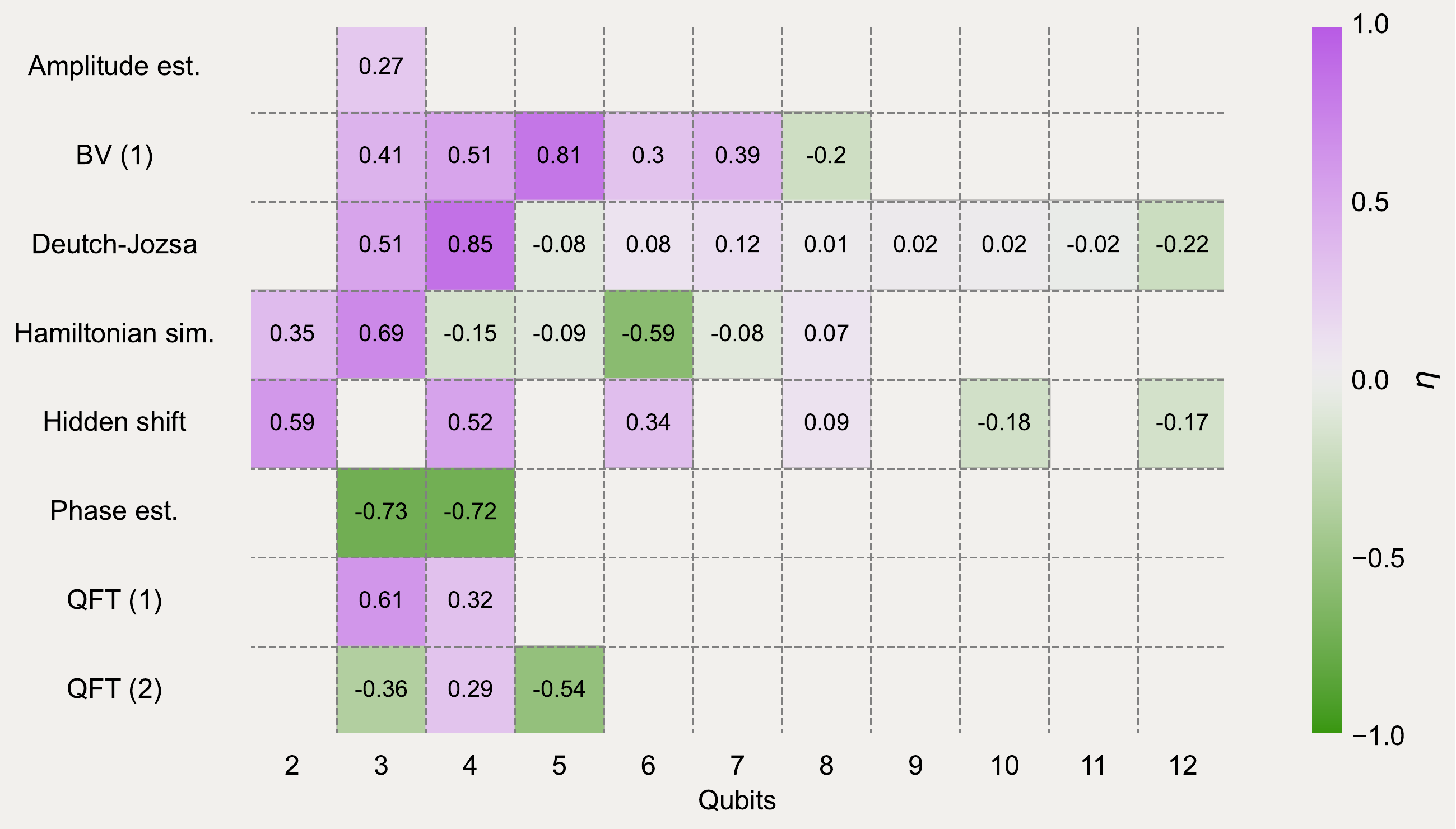}
\caption{\texttt{mapomatic} as a Qiskit transpiler pass versus PyTket.  The Sabre swap mapper was run 10 times for each Qiskit circuit with the circuit with the largest number of CNOT gates selected for execution.  All other execution parameters are the same as Fig.~(\ref{fig:qedc1}b).}
\label{fig:bad}
\end{figure}

\texttt{mapomatic} will remap any circuit passed to it that matches hardware topology and written in terms of the supported basis gates.  For compilation routines that contain stochastic components, such as the qubit routing methods in Qiskit, this means that the properties of the output circuits will follow a distribution of values.  For routing, this is a distribution in the total number of SWAP gates added to the circuit in order to satisfy the topology constraints of the target system.  Because each SWAP gate (equal to 3 CNOT gates) is costly to execute, there is a corresponding distribution in output fidelity when executing the same input circuit multiple times.  It can be the case that the variance in result fidelities is greater than the benefit of remapping, and optimizing routing before remapping with \texttt{mapomatic} is still an important part of the compilation workflow.  To overcome this, circuits should nominally be routed multiple times, which can be done in parallel, with the output circuit comprised of the fewest CNOT gates then passed on to \texttt{mapomatic}.  

To highlight the effect that routing has on the overall fidelity we will do the opposite of what is proposed above and compile circuits multiple times using Qiskit and the Sabre routing method, taking the one with the \textit{greatest} number of CNOT gates as the chosen circuit.  Figure (\ref{fig:bad}) shows the affect this has on the QED-C test circuits using the deterministic routing method in PyTket as the baseline result.  The detrimental consequence that added SWAP gates has on the resultant fidelity is clear when comparing against the results in Fig.~(\ref{fig:qedc1}b).

\bibliography{refs}
\end{document}